\documentclass[nofootinbib,showpacs,floatfix,onecolumn,prd]{revtex4-1}
\usepackage{graphicx,psfrag,amsmath,amssymb,amsfonts,latexsym,color,epsf,dcolumn,graphpap,braket}
\usepackage{caption,subcaption}
\usepackage{float}
\usepackage{lipsum}
\usepackage{enumerate}
\usepackage{cases}

\usepackage{hyperref}
\hypersetup{colorlinks=true,linktoc=page,linkcolor=blue,citecolor=red,urlcolor=cyan}

\usepackage{tikz-cd} 
\usepackage{tikz}
\usetikzlibrary{shapes.geometric, arrows}

\definecolor{red}{rgb}{1,0,0}
\definecolor{blue}{rgb}{0,0,1}
\definecolor{skyblue}{rgb}{0,0,.5}
\definecolor{green}{rgb}{0,1,0}
\definecolor{orange}{cmyk}{0,.4,1,0}

\usepackage[section]{placeins}

\usepackage{setspace}

\begin{document}
\onehalfspacing
\title{Vacuum polarization in the horizonless Bardeen metric}
\author{Andr\'es Boasso}
\author{Francisco D. Mazzitelli}
\affiliation{Centro At\'omico Bariloche,
Comisi\'on Nacional de Energ\'\i a At\'omica, 
R8402AGP Bariloche, Argentina}
\affiliation{Instituto Balseiro, 
Universidad Nacional de Cuyo,
Centro Atómico Bariloche, 
R8402AGP Bariloche, Argentina}

\begin{abstract}
We compute the renormalized stress-energy tensor for a massless quantum scalar field in the background of the horizonless Bardeen spacetime. Within the weak-field approximation, we show that the vacuum fluctuations differ significantly between conformally and nonconformally coupled fields, both in magnitude and in their behavior at short and intermediate distances. At large distances, we recover the universal asymptotic behavior previously observed in black hole and Newtonian star backgrounds. Going beyond the weak-field regime, we find that, for certain parameter ranges, the modes of the field can develop imaginary frequencies, leading to instabilities and an exponential growth of vacuum fluctuations. We also discuss critically the applicability of the anomaly-induced effective action for computing the renormalized stress-energy tensor in the conformally coupled case.
\end{abstract}

\date{24 November, 2025}


\maketitle
\section{Introduction}

The renormalized expectation value of the stress-energy tensor (RSET), $\langle T_{\mu\nu} \rangle$, plays a central role in quantum field theory in curved spacetime \cite{BooksQftcs}. It provides the key quantity for studying semiclassical effects such as Hawking radiation and the backreaction of quantum fields on the spacetime geometry. However, computing $\langle T_{\mu\nu} \rangle$ is notoriously difficult, even in highly symmetric settings, due to the nonlocal nature of quantum fluctuations and the need for renormalization.

Several approximation schemes have been developed to address this challenge, including weak-field expansions, dimensional reductions in spherically symmetric spacetimes, and fully numerical approaches \cite{ReviewRset}. Among the analytical methods, the covariant perturbation theory \cite{nonlocal}, based on an expansion in powers of the curvature, offers valuable conceptual insights, in regimes where the curvature is small. Although the covariant perturbation theory is not suitable for describing 
quantum effects around compact stars or black holes, 
it can still reveal important aspects of vacuum polarization.

In a previous work applied to Newtonian stars \cite{satz}, it was shown that, far from the source, the renormalized stress-energy tensor decays as $\langle T_{\mu\nu} \rangle \sim M/r^5$. Interestingly, the subleading terms in this expansion encode information about the internal structure of the star. This far-distance behavior was later shown \cite{Anderson2011} to be generic for static, spherically symmetric spacetimes that, for large $r$, are of the form
\begin{equation}
ds^2 = -\left(1-\frac{A}{r}\right) dt^2 +\left(1-\frac{B}{r}\right)dr^2 + r^2 d\Omega^2.
\end{equation}
Surprisingly, the leading behavior of the RSET at large distances is the same for Newtonian stars and for black holes, provided the quantum field is either minimally or conformally coupled \cite{universality}. These results rely on assuming the Boulware vacuum, which produces a RSET  regular for smooth stellar geometries but diverges at the horizon of black holes \cite{Christensen_Fulling}.

In stellar models without smooth enough matter profiles, the RSET can develop unphysical divergences at the stellar surface \cite{satz,jpnery}, reflecting the limitations of modeling the matter content as an idealized fluid concentrated in the star. To avoid such issues, in this work we focus on the horizonless Bardeen metric \cite{Bardeen}, a regular spacetime without curvature singularities or event horizons. In such backgrounds, the associated RSET should also remain regular everywhere.

We will compute $\langle T_{\mu\nu} \rangle$ analytically to leading order in the curvature expansion, and find quantitative differences between conformally and nonconformally coupled fields. Remarkably, for any value of the coupling, the asymptotic behavior at large $r$ coincides with that found in Newtonian stars, supporting the robustness of the $M/r^5$ decay pattern in vacuum polarization.

We will also explore a possible scenario where the Boulware vacuum may indeed become inadequate: the presence of tachyonic instabilities in the quantum field. Going beyond the weak-field approximation, we analyze the equation for the modes of the field and study the existence of bound states with imaginary frequencies $\omega^2 < 0$. When such modes are present, the RSET is expected to grow exponentially with time---a phenomenon previously described as the awakening of the vacuum \cite{awakening}.

Finally, we will discuss the results of a recent study~\cite{Boulware}, where the RSET for conformally coupled scalar fields in both \( 1+1 \) and \( 3+1 \) dimensions was computed using the effective action induced by the trace anomaly. In \( 3+1 \) dimensions, the authors numerically evaluated the RSET in a Bardeen-type, horizonless geometry, and found deviations from the behavior observed in Schwarzschild spacetime under the Boulware vacuum.  As we will argue, the discrepancy found in~\cite{Boulware} may stem from limitations of the anomaly-induced effective action \cite{Riegert}, which fails to capture the correct long-distance behavior of the RSET. 

This paper is organized as follows. In Sec.~\ref{m2:sec:bardeen} we review the basic aspects of the Bardeen metric. In Sec.~\ref{sec3} we compute the RSET in the Bardeen background using the covariant perturbative expansion method~\cite{nonlocal}. In this approach, the RSET is a nonlocal functional of the geometry; we retain terms linear in curvature and evaluate the result across the entire manifold. In the asymptotic region, our result confirms the previously observed universal decay ~\cite{universality}. In Sec.~\ref{sec4}, we revisit the definition of the Boulware vacuum in regular, horizonless geometries. We discuss the eventual emergence of vacuum instabilities beyond the weak-field regime, which may lead to an exponential growth of the vacuum polarization \cite{awakening}. In Sec.~\ref{sec5} we compare our results with those based on the anomaly-induced effective action. Finally, in Sec.~\ref{sec6} we summarize our findings.

\section{Bardeen metric}\label{m2:sec:bardeen}

The Bardeen spacetime~\cite{Bardeen} is a well-known example of a regular, static, spherically symmetric solution of Einstein’s equations. Its line element is given by  
\begin{equation}
\label{Bardeen_metric}
ds^2 = -f(r)\, dt^2 + \frac{dr^2}{f(r)} + r^2 \left( d\theta^2 + \sin^2\theta\, d\phi^2 \right),
\end{equation}
where the lapse function \( f(r) \) takes the form  
\begin{equation}
f(r) = 1 - \frac{2m(r)}{r}, \qquad m(r) = \frac{M r^3}{(R_0^2 + r^2)^{3/2}}\,.
\end{equation}
Here, \( M \) is a mass parameter and \( R_0 \) is a characteristic length scale.

This geometry arises as a solution of the Einstein equations coupled to a specific nonlinear electrodynamics in curved spacetime~\cite{beato}. In that context, \( R_0 \) is interpreted as the magnetic charge of a gravitating magnetic monopole. The geometry remains regular for all values of \( r \), and depending on the mass-to-charge ratio, different causal structures arise. There exists a critical value  
\[
M_c = \frac{3\sqrt{3}}{4} R_0,
\]
such that for \( M > M_c \), the spacetime describes a regular black hole with two horizons; for \( M < M_c \), it is completely regular and horizonless. The critical case \( M = M_c \) corresponds to an extremal black hole. In addition to its classical interpretation, the Bardeen geometry has also been considered as an effective model for quantum-corrected black holes~\cite{Ahkmedov}.

In the coordinates of Eq.~\eqref{Bardeen_metric}, the Ricci tensor takes the form
\begin{equation}
	R_{\mu}{}^{\nu} =
		\begin{pmatrix}
			\frac{3MR_0^2(2R_0^2-3r^2)}{(R_0^2 + r^2)^{7/2}} & 0 & 0 & 0 \\
			0 & \frac{3MR_0^2(2R_0^2-3r^2)}{(R_0^2 + r^2)^{7/2}} & 0 & 0  \\
			0 & 0 & \frac{6MR_0^2}{(R_0^2 + r^2)^{5/2}} & 0 \\
			0 & 0 & 0 & \frac{6MR_0^2}{(R_0^2 + r^2)^{5/2}} \\
		\end{pmatrix},
			\label{Bardeen_Rmunu}
\end{equation}
thus, the Ricci scalar is
\begin{equation}
	R = \frac{6MR_0^2(4R_0^2-r^2)}{(R_0^2 + r^2)^{7/2}}.
	\label{Bardeen_R}
\end{equation}
Observe that the components of the Ricci tensor are regular throughout the entire spacetime and remain nonzero near the origin. The temporal and radial components change sign at $r=\sqrt{2/3} R_0$, and all components decay as $1/r^5$ for large $r$. 

\section{The RSET in Bardeen spacetime: weak field approximation}\label{sec3}

In this Section, we attempt to compute the expectation value $\braket{T_{\mu\nu}}$ associated with the horizonless Bardeen-type metric. We consider the action of a massless scalar field $\phi$ in curved spacetime, given by
\begin{equation}
	S[\phi] = -\frac{1}{2}\int_{\mathcal{M}}\mathrm{d}^4x\sqrt{g}\left(\nabla_{\mu}\phi\nabla^{\mu}\phi + \xi R\phi^2\right),
\end{equation}
where $\mathcal{M}$ is a globally hyperbolic manifold and $\xi$ is the coupling to the curvature.  

The dynamics of the quantum field can be defined choosing a timelike Killing vector field $t^{\mu}$, which allows us to foliate the spacetime into a family of spacelike hypersurfaces $\Sigma_t$. A vacuum state can be defined as the state annihilated by all the operators $a_{\vec{k}}$ in the frequency modes expression of the field $\phi$,
\begin{equation}
	\phi(t,\vec{x}) = \int\mathrm{d}^{3}\vec{k}\,\left(a_{\vec{k}}u^{(+)}_{\vec{k}}(t,\vec{x}) + a_{\vec{k}}^{\dagger}u^{(-)}_{\vec{k}}(t,\vec{x})\right).
\end{equation}
Here, $\mathrm{d}^{3}\vec{k}$ represents the measure in the momentum space, defined to assure the commutation relations between the field and its conjugate momentum.

The modes $u^{(\pm)}_{\vec{k}}$ are solutions of the Klein-Gordon equation in curved spacetime identified by positive $(+)$ and negative $(-)$ frequencies, respectively. The choice of such modes is not unique, and it is possible to define a family of vacuum states, each corresponding to a different choice of the modes. For static metrics, positive and negative frequency can be defined with respect to the temporal Killing vector. This defines the Boulware vacuum, in which the modes can be factorized as
\begin{equation}
	u^{(+)}_{\vec{k}}(t,\vec{x}) = e^{-i\omega t}v_{\vec{k}}(\vec{x})\, , \quad  u^{(-)}_{\vec{k}}(t,\vec{x})=u^{(+)}_{\vec{k}}(t,\vec{x})^*
    \label{Boulware_modes}
\end{equation}

The RSET associated with a massless scalar field can be computed using a covariant perturbation theory. In four dimensions, and up to linear order in the curvature it is given by \cite{nonlocal,satz}
\begin{multline}
	\braket{T_{\mu}{}^{\nu}} = -\frac{1}{16\pi^{2}}\left(\xi-\frac{1}{6}\right)^{2}(\nabla_{\mu}\nabla^{\nu}-\delta_{\mu}{}^{\nu}\square) \log\left(\frac{-\square}{\mu^2}\right)R(x) \\
	- \frac{1}{5760\pi^{2}}\left[ (2\nabla_{\mu}\nabla^{\nu}+\delta_{\mu}{}^{\nu}\square) \log\left(\frac{-\square}{\mu^2}\right)R(x) - 6\square\log\left(\frac{-\square}{\mu^2}\right)R_{\mu}{}^{\nu}(x) \right].
	\label{Tmunu0}
\end{multline}
In this expression, the covariant derivatives and the d'Alembertian are taken as in flat spacetime.  

The action of the nonlocal operator $\log(-\square)$
on a time-independent  scalar test function $\varphi$ is given by \cite{satz}
\begin{equation}
	\log\left(\frac{-\square}{\mu^2}\right)\varphi(\vec{x}) = -\frac{1}{2\pi}\int\mathrm{d}^{3}\vec{x}'\, \frac{\varphi(\vec{x}')}{|\vec{x}-\vec{x}'|^{3}}\, ,
	\label{logbox}
\end{equation}
up to a local term proportional to $\varphi$.
Note that, in this approach, it is implicitly assumed that the quantum state of the field is the Boulware vacuum. 

To perform the calculations will work in spherical coordinates and split the problem into two: the computation of $\square\log(-\square)R_{\mu}{}^{\nu}$ and the computation of $\nabla_{\mu}\nabla^{\nu}\log(-\square)R$. 
It is worth remarking that, whenever the integral in Eq.~\eqref{logbox} is evaluated at a point $\vec x$ such that $\varphi(\vec x)\neq 0$, it has a logarithmic divergence that must be regularized. For Newtonian stars \cite{satz},  when the evaluation is performed outside the star, the integral in Eq.\eqref{logbox} is finite. 
This is not the case for the Bardeen metric. The isolation of the divergence is described below. From a physical point of view, the divergent term can be absorbed into the bare constant of the theory.

\subsection{Computation of $\square\log(-\square)R_{\mu}{}^{\nu}$}
As we are working to leading order in $M$,  Eq.~\eqref{Bardeen_Rmunu} is valid in {\it flat} spherical coordinates $x{}^{\mu}=\{t,r,\theta,\varphi\}$. We start by expressing $R_{\mu}{}^{\nu}$ in Cartesian coordinates $x'{}^{\mu}=\{t,x,y,z\}$ by using the equation
\begin{equation}
	R'_{\mu}{}^{\nu} = \frac{\partial x^{\rho}}{\partial x'^{\mu}}\frac{\partial x'^{\nu}}{\partial x^{\sigma}}R_{\rho}{}^{\sigma}.
\end{equation}

In Cartesian coordinates, we are able to compute $\log(-\square)R'_{\mu}{}^{\nu}$ component by component. The integrals in Eq.~\eqref{logbox} can be computed using spherical coordinates.\footnote{We stress that we are considering here the Cartesian components  $R'_{\mu}{}^{\nu}$, and  expressing each one in spherical coordinates.} Although $R'_{\mu}{}^{\nu}$ has several nonzero components, a simpler expression is obtained if we evaluate Eq.~\eqref{logbox} at some point on the $z$ axis,
\begin{equation}
    \log\left(\frac{-\square}{\mu^2}\right)R'_{\mu}{}^{\nu} = -\frac{1}{2\pi}\int_{0}^{\infty}\mathrm{d}r'\,r'{}^{2}\int_{0}^{\pi}\mathrm{d}\theta'\,\sin\theta'\int_{0}^{2\pi}\mathrm{d}\varphi'\,\frac{R'_{\mu}{}^{\nu}(r',\theta',\varphi')}{(r^{2}+r'{}^{2}-2rr'\cos\theta')^{3/2}},
    \label{logbox_Rmunu}
\end{equation}
and perform the $\varphi'$ integral,
\begin{equation}\begin{aligned}
	&\frac{1}{\pi}\int_{0}^{2\pi}\mathrm{d}\varphi'\,R'_{\mu}{}^{\nu}(r',\theta',\varphi') = \\
	&\begin{pmatrix}
		f_1(r') & 0 & 0 & 0 \\
		0 & f_2(r')+f_3(r')\cos^{2}\theta' & 0 & 0 \\
		0 & 0 & f_2(r')+f_3(r')\cos^{2}\theta' & 0 \\
		0 & 0 & 0 & f_4(r')-2f_3(r')\cos^{2}\theta' \\
	\end{pmatrix},
\end{aligned}\label{f_i0}\end{equation}
where
\begin{equation}
	\begin{aligned}
		f_1(r') &= MR_0^{2}\,\frac{6 \left(2 R_0^{2}-3 r'{}^{2}\right)}{\left(R_0^{2}+r'{}^{2}\right)^{7/2}} \\
		f_2(r') &= MR_0^{2}\,\frac{3 \left(4 R_0^{2}- r'{}^{2}\right)}{\left(R_0^{2}+r'{}^{2}\right)^{7/2}} \\
		f_3(r') &= MR_0^{2}\,\frac{15 r'{}^{2}}{\left(R_0^{2}+r'{}^{2}\right)^{7/2}} \\
		f_4(r') &= MR_0^{2}\,\frac{12}{\left(R_0^{2}+r'{}^{2}\right)^{5/2}}.
	\end{aligned}
	\label{f_i1}
\end{equation}

The integral over $\theta'$ can be performed using the change of variables $u=\cos\theta'$, which yields
\begin{align}
	g_{1}(r´) &:=\int_{-1}^{1}\mathrm{d}u\,(r^{2}+r'{}^{2}-2rr'\,u)^{-3/2} =\left\{
	\begin{aligned}
		&\frac{r^{2}}{(r-r')(r+r')} \quad&\text{ if }r'<r \\
		&\frac{2r^{3}}{r'(r-r')(r+r')} \quad&\text{ if }r'\geq r
	\end{aligned}\right.\label{eq:theta_int1}\\
	g_{2}(r´) &:=\int_{-1}^{1}u^2\mathrm{d}u\,(r^{2}+r'{}^{2}-2rr'\,u)^{-3/2} =\left\{
	\begin{aligned}
		&\frac{2(r^{2}+2r'{}^{2})}{3(r-r')(r+r')} \quad&\text{ if }r'<r \\
		&\frac{2r^{3}(r^{2}+2r'{}^{2})}{3r'{}^{3}(r-r')(r+r')} \quad&\text{ if }r'\geq r,
	\end{aligned}\right.\label{eq:theta_int2}
\end{align}
thus, replacing Eqs.~\eqref{f_i0}--\eqref{eq:theta_int2} in \eqref{logbox_Rmunu}, we obtain
\begin{align}
	&\log\left(\frac{-\square}{\mu^2}\right) R'_{\mu}{}^{\nu} =\notag\\
	&-\frac{1}{2}\begin{pmatrix}
		\int_{0}^{\infty}\mathrm{d}r'{r'}^{2}f_1\,g_1 & 0 & 0 & 0 \\
		0 & \int_{0}^{\infty}\mathrm{d}r'{r'}^{2}(f_2\,g_1+f_3\,g_2) & 0 & 0 \\
		0 & 0 & \int_{0}^{\infty}\mathrm{d}r'{r'}^{2}(f_2\,g_1+f_3\,g_2) & 0 \\
		0 & 0 & 0 & \int_{0}^{\infty}\mathrm{d}r'{r'}^{2}(f_4\,g_1-2f_3\,g_2) \\
	\end{pmatrix},
\end{align}

The last integral in $r'$ requires regularization in all the cases: notice that all the divergent contributions arise from the factor $(r-r')^{-1}$ present in each case of Eqs.~\eqref{eq:theta_int1} and \eqref{eq:theta_int2}. In order to identify the divergent terms, we split each integral into two intervals $(0,r-\Delta)\cup(r+\Delta,\infty)$ and take the limit $\Delta\rightarrow 0^{+}$, obtaining the following:
\begin{align}
	&\log\left(\frac{-\square}{\mu^2}\right) R'_{\mu}{}^{\nu} =\notag \\
	&\begin{pmatrix}
		F_1+f_1\log\left(\mu\Delta\right) & 0 & 0 & 0 \\
		0 & F_2+F_3+(f_2+f_3)\log\left(\mu\Delta\right) & 0 & 0 \\
		0 & 0 & F_2+F_3+(f_2+f_3)\log\left(\mu\Delta\right) & 0 \\
		0 & 0 & 0 & F_4-2F_3+(f_4-2f_3)\log\left(\mu\Delta\right) \\
	\end{pmatrix},
\label{F_i0}
\end{align}
where
\begin{equation}
	\begin{aligned}
		F_1(r) &= M\,\frac{2 \left(17 R_0^4-28 R_0^2 r^2+r^4\right)}{\left(R_0^2+r^2\right)^{7/2}} + f_1(r)\log\left(\frac{R_0}{\mu(R_0^2 + r^2)}\right)\\
		F_2(r) &= M\,\frac{31R_0^4 - 16R_0^2r^2 - r^4}{\left(R_0^2+r^2\right)^{7/2}} + f_2(r)\log\left(\frac{R_0}{\mu(R_0^2 + r^2)}\right)\\
		F_3(r) &= M\,\frac{-R_0^4 + 44R_0^2r^2 - r^4}{\left(R_0^2+r^2\right)^{7/2}} + f_3(r)\log\left(\frac{R_0}{\mu(R_0^2 + r^2)}\right)\\
		F_4(r) &= M\,\frac{4 \left(7 R_0^2 -r^2\right)}{\left(R_0^2+r^2\right)^{5/2}} + f_4(r)\log\left(\frac{R_0}{\mu(R_0^2 + r^2)}\right).
	\end{aligned}
	\label{F_i1}
\end{equation}
The procedure we followed is equivalent to the regularization of a generalized function, studied in detail in \cite{Gelfand}.

The expression in Eq.~\eqref{F_i0} was obtained in Cartesian coordinates and is valid only at any point on the $z$ axis. Since we know that the Ricci tensor is spherically symmetric, we can extend this result to be valid off the $z$ axis via a generic spatial rotation. Observe that, for any spherically symmetric tensor that in Cartesian coordinates and on the $z$ axis has the form $W'{}_{\mu}{}^{\nu} = \text{diag}(a,c,c,b)$, its expression in spherical coordinates is $W{}_{\mu}{}^{\nu} = \text{diag}(a,b,c,c)$, and the latter is valid in the whole space. Therefore, Eq.~\eqref{F_i0} becomes
\begin{equation}
	\log\left(\frac{-\square}{\mu^2}\right) R_{\mu}{}^{\nu}(r) = \\
	\begin{pmatrix}
		F_1(r) & 0 & 0 & 0 \\
		0 & F_4(r)-2F_3(r) & 0 & 0 \\
		0 & 0 & F_2(r)+F_3(r) & 0 \\
		0 & 0 & 0 & F_2(r)+F_3(r)
	\end{pmatrix}+R_{\mu}{}^{\nu}(r)\log\left(\mu\Delta\right).
\label{logboxRmunu}
\end{equation}

We can drop the local, divergent term proportional to $\log(\mu\Delta)$ in Eq.~\eqref{logboxRmunu}. Physically, we are not discarding a divergent contribution, but we are absorbing it into the bare constants of the theory. The length scale $1/\mu$ can be chosen to equal $R_0$ in order not to introduce new dimensional parameters into our results. However, it should be kept arbitrary if one is interested, for example, in studying the renormalization-group flow of the dressed constants.


Finally, we compute $\square\log\left(-\square\right) R_{\mu}{}^{\nu}$ by acting with the flat d'Alembertian operator in spherical coordinates on Eq.~\eqref{logboxRmunu}. We do not display the result as its components involve lengthy and cumbersome expressions.

\subsection{Computation of $\nabla_{\mu}\nabla^{\nu}\log(-\square)R$}
Fortunately, no further complicated computations are needed. Since we know $R(r)$ (Eq.~\eqref{Bardeen_R}), we can relate its expression to the ones presented in Eq.~\eqref{f_i1} via $R(r)=2f_2(r)$; therefore, adding a factor of $2$ due to the integral on Eq.~\eqref{f_i0}, we get
\begin{equation}
	\log\left(\frac{-\square}{\mu^2}\right) R(r) = 4 F_2(r).
	\label{logboxR}
\end{equation}

In a similar way we obtain
\begin{equation}
	\nabla_{\mu}\nabla^{\nu}\log\left(\frac{-\square}{\mu^2}\right)R = 4
	\begin{pmatrix}
		0 & 0 & 0 & 0 \\
		0 & F_2'' & 0 & 0 \\
		0 & 0 & \frac{F_2'}{r} & 0 \\
		0 & 0 & 0 & \frac{F_2'}{r}
	\end{pmatrix} ,
\end{equation}
and therefore
\begin{equation}
	\square\log\left(\frac{-\square}{\mu^2}\right) R = 4 \left(F_2'' +2\frac{F_2'}{r}\right).
	\label{boxlogboxR}
\end{equation}

Replacing  the obtained results in Eq.~\eqref{Tmunu0}, we can write down $\braket{T_{\mu}{}^{\nu}}$ as a sum of two contributions,
\begin{equation}
	\braket{T_{\mu}{}^{\nu}} = \left(\xi-\frac{1}{6}\right)^2\braket{T_{\mu}{}^{\nu}}^{(\text{nc})} + \braket{T_{\mu}{}^{\nu}}^{(\text{c})},
	\label{final_Tmunu}
\end{equation}
where
\begin{equation}\scriptsize\begin{aligned}
	&\braket{T_{t}{}^{t}}^{(\text{nc})} = -\frac{M}{\pi^2}\frac{6 r^6+337 r^4 R_0^2-2228 r^2 R_0^4+819 R_0^6}{4 \left(r^2+R_0^2\right)^{11/2}} - \frac{M}{\pi^2}\frac{15 \left(4 r^4 R_0^2-41 r^2 R_0^4+18 R_0^6\right)}{4 \left(r^2+R_0^2\right)^{11/2}}\log\left(\frac{R_0^2}{R_0^2 + r^2}\right) \\
	&\braket{T_{r}{}^{r}}^{(\text{nc})} = \frac{M}{\pi^2}\frac{3 r^4+82 r^2 R_0^2-273 R_0^4}{2 \left(r^2+R_0^2\right)^{9/2}} +\frac{M}{\pi^2}\frac{15 R_0^2 \left(r^2-6 R_0^2\right)}{2 \left(r^2+R_0^2\right)^{9/2}}\log\left(\frac{R_0^2}{R_0^2 + r^2}\right) \\
	&\braket{T_{\theta}{}^{\theta}}^{(\text{nc})} = \braket{T_{\varphi}{}^{\varphi}}^{(\text{nc})} = -\frac{M}{\pi^2}\frac{9 r^6+422 r^4 R_0^2-2419 r^2 R_0^4+546 R_0^6}{4 \left(r^2+R_0^2\right)^{11/2}} - \frac{M}{\pi^2}\frac{15 \left(5 r^4 R_0^2-46 r^2 R_0^4+12 R_0^6\right)}{4 \left(r^2+R_0^2\right)^{11/2}} \log\left(\frac{R_0^2}{R_0^2 + r^2}\right),
\end{aligned}\end{equation}
and
\begin{equation}\scriptsize\begin{aligned}
	&\braket{T_{t}{}^{t}}^{(\text{c})} = \frac{M}{\pi^2}\frac{6 r^6-485 r^4 R_0^2+982 r^2 R_0^4-216 R_0^6}{360 \left(r^2+R_0^2\right)^{11/2}} - \frac{M}{\pi^2}\frac{16 r^4 R_0^2-38 r^2 R_0^4+9 R_0^6}{48 \left(r^2+R_0^2\right)^{11/2}}\log\left(\frac{R_0^2}{R_0^2 + r^2}\right) \\
	&\braket{T_{r}{}^{r}}^{(\text{c})} = \frac{M}{\pi^2}\frac{3 r^4-101 r^2 R_0^2+72 R_0^4}{360 \left(r^2+R_0^2\right)^{9/2}} +\frac{M}{\pi^2}\frac{-4 r^2 R_0^2 + 3 R_0^4}{48 (r^2 + R_0^2)^{9/2}}\log\left(\frac{R_0^2}{R_0^2 + r^2}\right) \\
	&\braket{T_{\theta}{}^{\theta}}^{(\text{c})} = \braket{T_{\varphi}{}^{\varphi}}^{(\text{c})} = \frac{M}{\pi^2}\frac{-9 r^6+583 r^4 R_0^2-953 r^2 R_0^4+144 R_0^6}{720 \left(r^2+R_0^2\right)^{11/2}} + \frac{M}{\pi^2}\frac{20 r^4 R_0^2-37 r^2 R_0^4+6 R_0^6}{96 \left(r^2+R_0^2\right)^{11/2}} \log\left(\frac{R_0^2}{R_0^2 + r^2}\right).
\end{aligned}\end{equation}

\begin{figure}[htbp]
    \centering
    \begin{subfigure}[t]{0.48\textwidth}
        \centering
        \includegraphics[width=\textwidth]
        {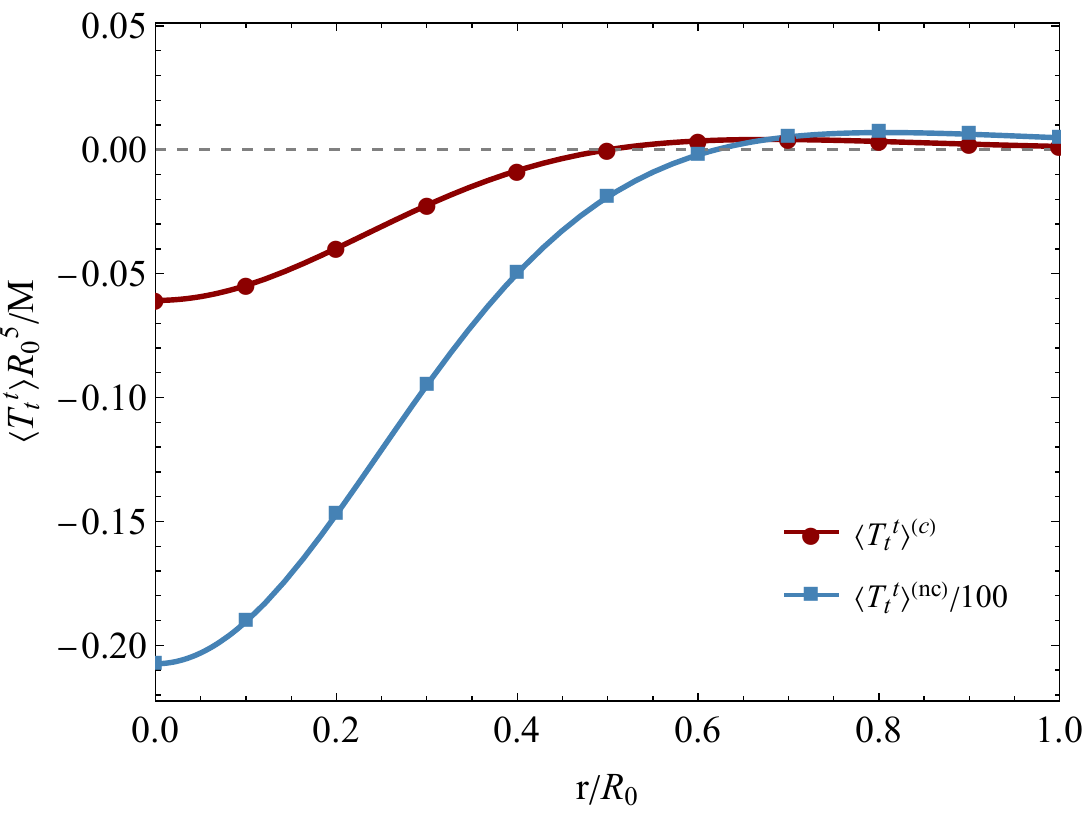}        
        \caption{}
        \label{fig:subfig1}
    \end{subfigure}
    \hfill
    \begin{subfigure}[t]{0.48\textwidth}
        \centering
        \raisebox{-1mm}{\includegraphics[width=\textwidth]{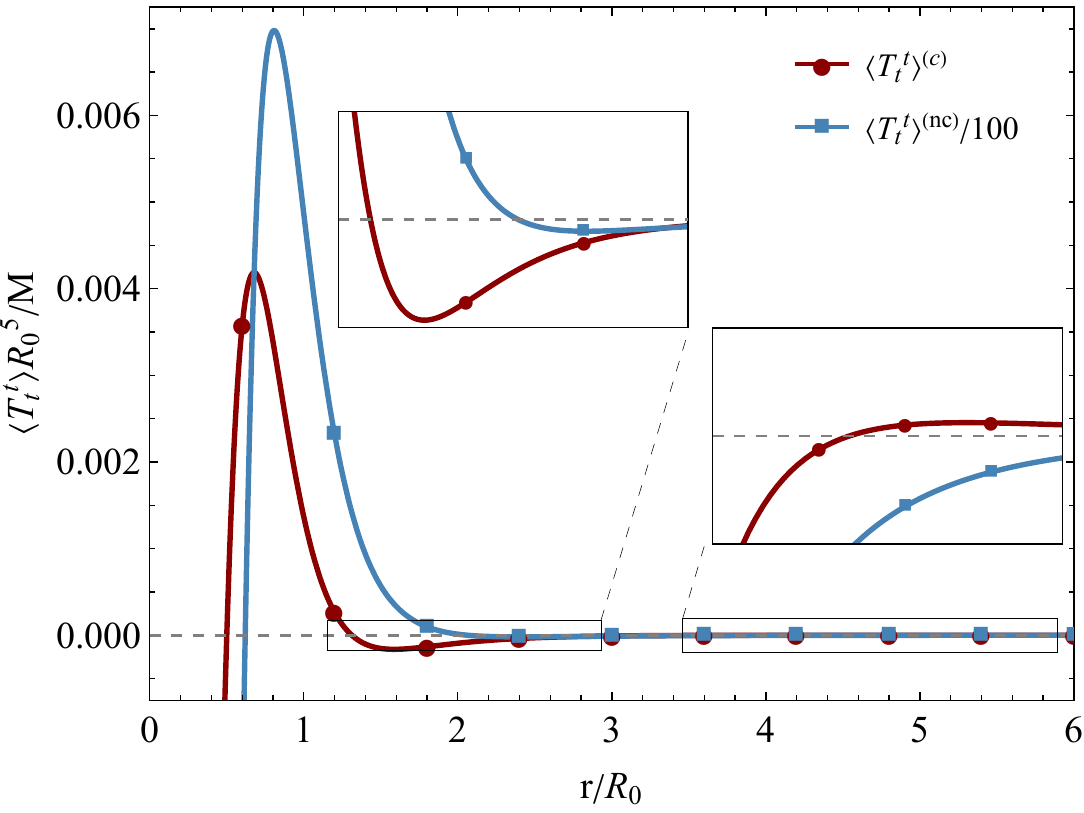}}      
        \caption{}
        \label{fig:subfig2}
    \end{subfigure}
    \caption{\centering Energy density associated with the nonconformal and conformal parts of the stress-energy tensor, $\braket{T_{t}{}^{t}}^{(\text{nc})}$ and $\braket{T_{t}{}^{t}}^{(\text{c})}$, as functions of the distance from the origin. The quantity $\braket{T_{t}{}^{t}}^{(\text{nc})}$ has been rescaled by a factor of $1/100$ in order to facilitate comparison with $\braket{T_{t}{}^{t}}^{(\text{c})}$. Figure~\ref{fig:subfig1} displays the general behavior of the energy densities. The zoom in Fig.~\ref{fig:subfig2} shows the structure for $r\simeq R_0$. At large $r$, both energy densities fall off as $\sim 1/r^5$ [see Eq.~\eqref{larger}].}
    \label{fig:energy}
\end{figure}

Figure~\ref{fig:energy} shows the energy density corresponding to both the conformal and nonconformal components of the stress tensor. As displayed in Fig.~\ref{fig:subfig1}, the nonconformal part is approximately two orders of magnitude larger than the conformal one near the origin.
The densities have a rich structure for $r\simeq R_0$, shown in Fig.~\ref{fig:subfig2}: note that the number of zero crossings differs between the two curves. Finally, both components exhibit the expected asymptotic behavior for large values of $r$, according to Eq.~\eqref{larger} below.

As expected, the tensor $\braket{T_{\mu}{}^{\nu}}^{(\text{c})}$ is traceless, since there is no trace anomaly up to first order in the weak-field approximation. Additionally, we can check that the tensor $\braket{T_{\mu}{}^{\nu}}$ in Eq.~\eqref{final_Tmunu} satisfies the conservation law for any value of $\xi$. Indeed, in these coordinates, the conservation law reduces to a single nontrivial equation,
\begin{equation}
	\frac{2\left(\braket{T_r{}^r}-\braket{T_{\theta}{}^{\theta}}\right)}{r} + \partial_r\braket{T_r{}^r}=0.
\end{equation}

Finally, we consider a series in powers of $1/r$ to identify the leading order in Eq.~\eqref{final_Tmunu},

\begin{equation}\begin{aligned}\label{larger}
	\braket{T_t{}^t} &= -\frac{3M}{2\pi^2}\left(\xi-\frac{1}{6}\right)^2\frac{1}{r^5}+\frac{M}{60\pi^2r^5} \\
	\braket{T_r{}^r} &= \frac{3M}{2\pi^2}\left(\xi-\frac{1}{6}\right)^2\frac{1}{r^5}+\frac{M}{120\pi^2r^5} \\
	\braket{T_{\theta}{}^{\theta}} &= \braket{T_{\varphi}{}^{\varphi}} = -\frac{9M}{4\pi^2}\left(\xi-\frac{1}{6}\right)^2\frac{1}{r^5}-\frac{M}{80\pi^2r^5}.
\end{aligned}\end{equation}

It is worth remarking that, in the long-distance regime, these results coincide with those of a Newtonian star, for an arbitrary coupling to the curvature. This is remarkable, since the Ricci scalar vanishes outside Newtonian stars and is everywhere different from zero in the Bardeen metric. Moreover, for minimal and conformal couplings the leading behavior coincides with the result for Schwarszchild metric computed using the Boulware vacuum, extending the universal behavior found in Ref.~\cite{universality}. On the other hand, our results for the leading order of the RSET in Eq.~\eqref{larger} are consistent with those reported in Ref.~\cite{Anderson2011}, which were obtained using a different method where the choice of the Boulware vacuum state is explicit. This agreement is to be expected, because of the implicit choice of the Boulware vacuum in the nonlocal effective action approach.

At short and intermediate distances, the behavior in the Bardeen spacetime differs from that of a Newtonian star, and depends of course on its inner structure. For instance, for a Newtonian, constant-density star of mass $M$ and radius $R_0$,  the RSET is given by \cite{satz},

\begin{equation}
\begin{aligned}
\left\langle T_{t}{}^{t} \right\rangle &=
\, \frac{3}{4\pi^2}\frac{M}{R_0^3}
\left[
\left( \xi - \frac{1}{6} \right)^{2}
- \frac{1}{90}
\right]
\frac{3R_{0}^{2} - r^{2}}{(R_{0}^{2} - r^{2})^{2}},
\\[1em]
\left\langle T_{r}{}^{r} \right\rangle &=
\,\frac{3}{2\pi^2}\frac{M}{R_0^3}
\left[
\left( \xi - \frac{1}{6} \right)^{2}
+ \frac{1}{180}
\right]
\frac{1}{R_{0}^{2} - r^{2}},
\\[1em]
\left\langle T_{\theta}{}^{\theta} \right\rangle &=
\,\frac{3}{2\pi^2}\frac{M}{R_0^3}
\left[
\left( \xi - \frac{1}{6} \right)^{2}
+ \frac{1}{180}
\right]
\frac{R_{0}^{2}}{(R_{0}^{2} - r^{2})^{2}},
\end{aligned}
\end{equation}
for $r<R_0$ and 
\begin{equation}
\begin{aligned}
\left\langle T_{t}{}^{t} \right\rangle &=
-\,\frac{3}{4\pi^2}\frac{M}{R_0^3}\left[
\left( \xi - \frac{1}{6} \right)^{2}
- \frac{1}{90}
\right]
\frac{2 R_{0}^{3}}{r (r^{2} - R_{0}^{2})^{2}},
\\[1em]
\left\langle T_{r}{}^{r} \right\rangle &=
\,\frac{3}{2\pi^2}\frac{M}{R_0^3}
\left[
\left( \xi - \frac{1}{6} \right)^{2}
+ \frac{1}{180}
\right]
\frac{R_{0}^{3}}{r^{3} (r^{2} - R_{0}^{2})},
\\[1em]
\left\langle T_{\theta}{}^{\theta} \right\rangle &=
-\,\frac{3}{2\pi^2}\frac{M}{R_0^3}
\left[
\left( \xi - \frac{1}{6} \right)^{2}
+ \frac{1}{180}
\right]
\frac{R_{0}^{3} (3r^{2} - R_{0}^{2})}{2 r^{3} (r^{2} - R_{0}^{2})^{2}},
\end{aligned}
\end{equation}
for $r>R_0$. We see that  for this Newtonian  star the RSET diverges at the surface, while in the Bardeen spacetime it stays regular everywhere (see Fig.~2).


\begin{figure}[htbp]
    \centering
    \includegraphics[width=0.48\textwidth]{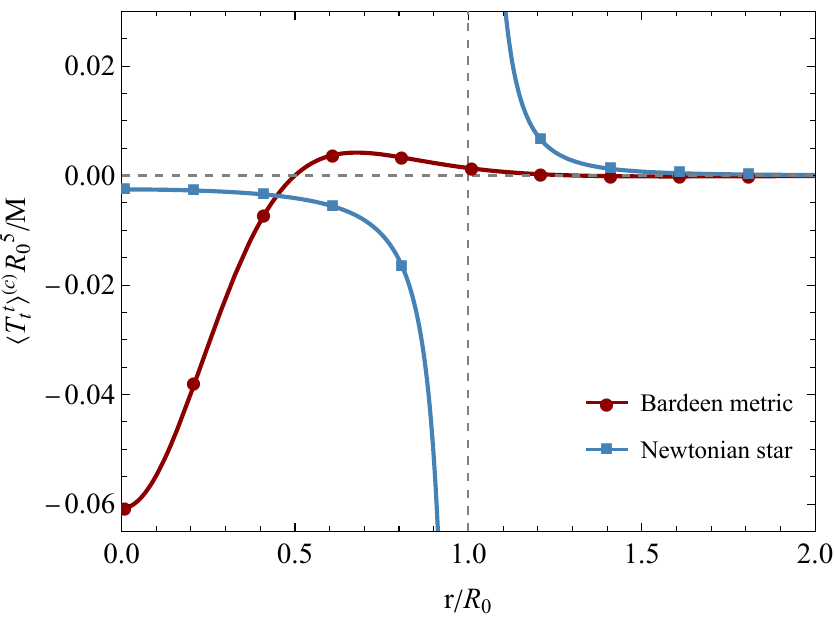}
    \caption{\centering Comparison between the component $\langle T_t{}^t\rangle$ associated with the Bardeen spacetime and the corresponding component for a Newtonian, pressureless star of constant density. We use the same parameters $M$ and $R_0$ for both geometries.   The plot corresponds to the conformal case $\xi=1/6$.}
    \label{fig:star}
\end{figure}


\section{Beyond the weak field approximation: Boulware vacuum and eventual tachyonic instabilities}\label{sec4}

 In a series of works \cite{universality}, it has been shown that sufficiently compact objects can trigger an exponential growth in the vacuum energy of quantum scalar fields. This phenomenon arises due to tachyonic instabilities, where the Klein-Gordon equation admits modes with negative squared frequencies. In such cases, the scalar field lacks a true ground state, and the usual construction of the Boulware vacuum becomes invalid. This so-called ``awakening of the vacuum'' may eventually lead to interesting astrophysical consequences, as discussed in Ref.\cite{awakening}. For example, neutron stars with a given mass and radius would not be stable in the presence of a quantum scalar field with coupling to the curvature such that there are tachyonic instabilities. The observation of such stars could be used to rule out the  existence of scalar fields with those couplings.

In this Section we will discuss whether the horizonless Bardeen metric may induce tachyonic instabilities or not for the quantum scalar field. We will demonstrate that, in the weak-field approximation, the Boulware vacuum is a stable quantum state. However, as we will see, the Boulware vacuum may not be stable for strong gravitational fields, due to the presence of a tachyonic instabilities. Although there is no explicit tachyonic term $m^2<0$ in the action, the effective mass term can be negative due to the nontrivial geometry. To see this, we will analyze the equations that satisfy the field modes $u^{(\pm)}_{\vec{k}}$, looking for values of $\vec{k}$ such that $\omega$ is imaginary, i.e., for unstable modes that increase their amplitude exponentially in time.

The modes of a scalar massless field in the Bardeen metric are solutions for the Klein-Gordon equation
\begin{equation}
	\left(\square-\xi R\right)u^{(\pm)}_{\vec{k}}(t,\vec{x}) = 0,
	\label{KG}
\end{equation}
where the functions $u^{(\pm)}_{\vec{k}}$ are factorized as shown in Eq.~\eqref{Boulware_modes}. Since the Bardeen metric is spherically symmetric, we propose
\begin{equation}
	u^{(\pm)}_{\omega l m} = \frac{e^{\mp i\omega t}}{\sqrt{2\omega}}\,\frac{F_{\omega l}(r)}{r} \, Y_{l}^{m}(\theta,\varphi),
	\label{modes0}
\end{equation}
where $Y_{l}^{m}(\theta,\varphi)$ is the spherical harmonic of order $(l,m)$, and the indices $l,m,\omega$ play the role of $\vec{k}$ in Eq.~\eqref{KG}. Using the operator $\square=\nabla^{\mu}\nabla_{\mu}$ associated to the metric Eq.~\eqref{Bardeen_metric} in Eq.~\eqref{KG} yields
\begin{equation}
	\frac{1}{r^2}\frac{\partial}{\partial r}\left(r^2f\frac{\partial u^{(\pm)}_{\omega l m}}{\partial r}\right) + \frac{1}{r^2\sin\theta}\frac{\partial}{\partial \theta}\left(\sin\theta\frac{\partial u^{(\pm)}_{\omega l m}}{\partial \theta}\right)+\frac{1}{r^2\sin^2\theta}\frac{\partial^2 u^{(\pm)}_{\omega l m}}{\partial \theta^2} -\xi R u^{(\pm)}_{\omega l m}= \frac{1}{f}\frac{\partial^2 u^{(\pm)}_{\omega l m}}{\partial t^2}.
	\label{modes1}
\end{equation}

Now, we consider a change of the radial variable defined by $\mathrm{d}r/\mathrm{d}\chi = f$, and therefore, replacing Eq.~\eqref{modes0} in Eq.~\eqref{modes1} yields
\begin{equation}
	-\frac{\mathrm{d}^2F_{\omega l}}{\mathrm{d}\chi^2} + \underbrace{f\left(\xi R + \frac{l(l+1)}{r^2} + \frac{f'}{r}\right)}_{=V_{\text{eff}}^{(l)}}F_{\omega l} = \omega^2 F_{\omega l},
	\label{modes2}
\end{equation}
where $f'=\mathrm{d}f/\mathrm{d}r$, and $r=r(\chi)$. The spectrum of eigenvalues $\omega^2$ in Eq.~\eqref{modes2} determines if a tachyonic instability is present in the theory. If Eq.~\eqref{modes2} admits solutions with $\omega^2<0$, then $\text{Im}(\omega)\neq 0$, and the amplitude of some modes grows exponentially in time. We stress that for $\omega^2>0$ the solutions $F_{\omega l}$ are fully determined by the condition of being regular at the origin and by the normalization of the modes. However, when $\omega^2 < 0$ the solutions must vanish as $r\to\infty$.

In order to analyze the existence of solutions with $\omega^2<0$, we will exploit its similarity to the one-dimensional Schrödinger equation with potential $V_{\text{eff}}^{(l)}$. If we consider another potential $\tilde V_{\text{eff}}$ such that $\tilde V_{\text{eff}}(\chi)<V_{\text{eff}}^{(l)}(\chi)\:\:\forall\chi$, and the spectrum of eigenvalues of Eq.~\eqref{modes2} associated with $\tilde V_{\text{eff}}$ is bounded from below, then ${\tilde\omega_0}^2<\omega_0^2$, where $\omega_0^2$ and ${\tilde\omega_0}^2$ are the lowest eigenvalues of Eq.~\eqref{modes2} for the potentials $V_{\text{eff}}^{(l)}$ and $\tilde V_{\text{eff}}$, respectively. Using this property,  we will find constraints on the parameters $M$, $R_0$, and $\xi$ that ensure nonnegative eigenvalues. 

We first notice that $V_{\text{eff}}^{(l=0)}<V_{\text{eff}}^{(l\geq1)}$; therefore, any $\tilde V_{\text{eff}}$ such that $\tilde V_{\text{eff}}<V_{\text{eff}}^{(l=0)}$ will cover all cases. For simplicity, we denote $V_{\text{eff}}^{(l=0)}\equiv V_{\text{eff}}$, whose explicit form is
\begin{equation}
	V_{\text{eff}} = \frac{2 M \left(\left(R_0^2+r^2\right)^{3/2}-2 M r^2\right) \left(2 (6 \xi -1) R_0^4 -(3 \xi +1)R_0^2 \, r^2+r^4\right)}{\left(R_0^2+r^2\right)^5}.
	\label{Veff0}
\end{equation}

The simplest case is to take $\tilde V_{\text{eff}}=0$, since Eq.~\eqref{modes2} with this potential does not admit negative eigenvalues, and therefore $\omega^2\geq0$. In this case, the constraint reads
\begin{equation}
	0.225\approx\frac{7-2\sqrt{10}}{3}\leq\xi\leq\frac{7+2\sqrt{10}}{3}\approx4.442\, .
\end{equation}
However, this is a strong constraint that excludes both the minimally ($\xi=0$) and conformally coupled ($\xi=1/6$) cases.

\begin{figure}[htbp]
  \centering
	\includegraphics[width=0.48\textwidth]{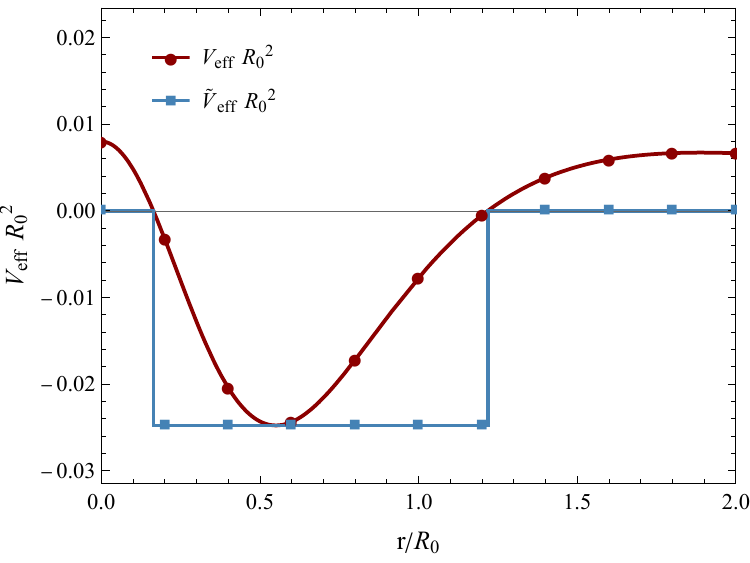}
  \caption{\centering Effective potential of Eq.~\eqref{Veff0}, $\xi=0.17$ and $M/R_0=0.1$ (red curve with circle markers), and the proposed function $\tilde V_{\text{eff}}$ as a square potential well (blue curve with square markers).}
  \label{fig:square}
\end{figure}

We would like to relax this restriction to include the minimal and conformally coupled cases and, ideally, guarantee the absence of tachyonic modes in the weak-field limit.
We can adopt $\tilde V_{\text{eff}}$ as a square potential well (see Fig.~\ref{fig:square}). The existence of bound states (negative eigenvalue) in this case can be easily determined: if $L$ is the width of the well and $V_0$ its depth, there exist bounded states with Dirichlet conditions at $r=0$ and $r\rightarrow\infty$ if and only if $V_0L^2>\pi^2/4$. 

The width $L$ of the well must be evaluated in the $\chi$ coordinate. As we do not have an analytical formula that relates $\chi$ and $r$, we look for a conservative constraint. The differential relationship between $\chi$ and $r$ is $\mathrm{d}\chi = \mathrm{d}r/f$, and the function $f$ satisfies $f(r)\geq1-4M/3\sqrt{3}R_0>0$. Therefore, an interval in the $\chi$ coordinate satisfies
\begin{equation}
	L = \chi_1-\chi_0 = \int_{r_0}^{r_1}\frac{\mathrm{d}r}{f(r)} \leq \frac{1}{1-\frac{4M}{3\sqrt{3}R_0}}(r_1-r_0)\equiv L_0.
	\label{Lbound}
\end{equation}
The interval $[r_0,r_1]$ where the effective potential in Eq.~\eqref{Veff0} remains negative can be analytically found, so Eq.~\eqref{Lbound} shows an upper bound to the width of the well. 
Given a combination of parameters \( \xi \), \( M \), and \( R_0 \), if the condition 
\( V_0 L^2 \leq V_0 L_0^2 < \pi^2 / 4 \) is satisfied, then there are no negative eigenvalues 
associated with \( \tilde{V}_{\text{eff}} \), and consequently, none associated with 
\( V_{\text{eff}} \) either.

The location of the minimum of \( V_{\text{eff}} \) is determined by solving a seventh-degree polynomial equation, which must be done numerically. Both the width and depth of the potential well depend solely on the coupling \( \xi \) and the ratio \( M / R_0 \). We recall that, in this metric, the weak-field limit corresponds to \( M \ll R_0 \).

\begin{figure}[htbp]
  \centering
	\includegraphics[width=0.48\textwidth]{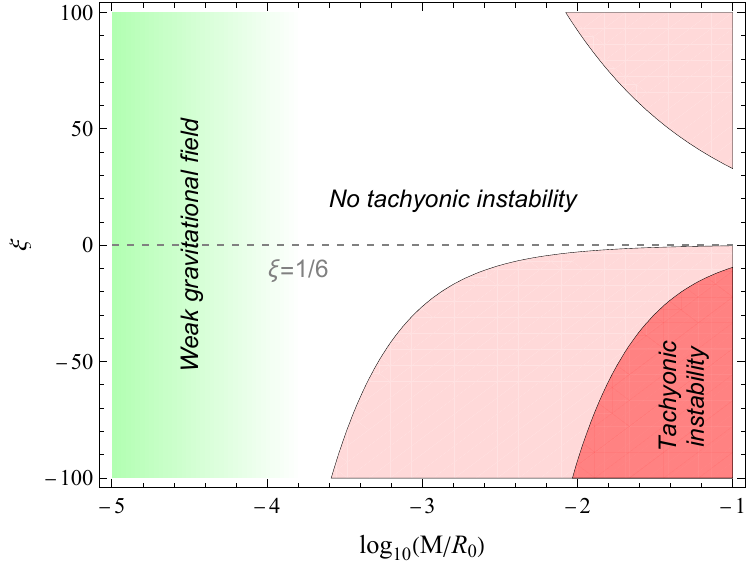}
  \caption{\centering Certain combinations of the parameters \( \xi \), \( M \), and \( R_0 \) lead to a tachyonic instability (dark red-shaded region). In the light red-shaded region, neither the presence of a tachyonic instability nor the existence of a stable Boulware vacuum can be guaranteed. Outside the red-shaded regions, no tachyonic instability occurs. The dashed horizontal line corresponds to the conformal coupling case \( \xi = 1/6 \). In the weak-field regime \( M \ll R_0 \) (left side of the plot), no tachyonic instability is present for any values of \( \xi \), \( M \), and \( R_0 \).
}
  \label{fig:zone1}
\end{figure}

We show the results of this analysis in Fig.~\ref{fig:zone1}. The red-shaded zones represent the combinations of parameters $\xi$ and $M/R_0$ whose associated square potential well presents at least one negative eigenvalue, indicating a possible presence of tachyonic instability in the original effective potential. Conversely, the white regions correspond to parameter values where the square well does not admit negative eigenvalues, ensuring the stability of the Boulware vacuum. As can be seen, for sufficiently small $M/R_0$ (the weak-field regime), any value of the coupling $\xi$ lies within the stable region, confirming that the Boulware vacuum remains stable in this limit.  

Using similar reasoning, one can identify regions in parameter space where negative eigenvalues arise. In particular, it is possible to construct a square potential well such that $V_{\text{eff}} > V_{\text{eff}}^{(l=0)}$. The existence of a bound state with a negative eigenvalue in this effective potential ensures the presence of tachyonic instabilities in the Bardeen metric (see the dark red-shaded region in Fig.~\ref{fig:zone1}).  This provides a new example of the interesting phenomenon of vacuum tachyonic instability (see Ref.\cite{Vanzella} for previous examples).

\section{Anomaly-induced RSET}\label{sec5}

Recently, Ref.~\cite{Boulware} presented a numerical computation of the RSET in Bardeen-type spacetimes for the conformal coupling \( \xi = 1/6 \), using a method based on the anomaly-induced effective action and focusing on the strong-field case \( M = R_0/2 \). Their analysis found an asymptotic decay of the RSET components as \( 1/r^3 \), leading the authors to suggest that the Boulware vacuum might not be suitable for this background. For general parameters \( M \) and \( R_0 \), such a \( 1/r^3 \) behavior could, in principle, arise through inverse powers of \( R_0 \), for instance as \( M^2/(R_0^3 r^3) \) or \( M^3/(R_0^4 r^3) \).

Our results, however, indicate a different asymptotic behavior in the weak-field regime. Indeed, for \( M \gg R_0 \), the components of the RSET decay as \( M/r^5 \), i.e., independently of the value of \( R_0 \). This is consistent with the more general analysis of Ref.~\cite{Anderson2011}, which adopts the Boulware vacuum without assuming weak gravity. Moreover, in the covariant perturbative approach employed here—originally developed in Euclidean spacetime~\cite{nonlocal}—the Wick rotation naturally selects the Boulware vacuum in static backgrounds. These considerations suggest that the Weyl-invariant part of the effective action plays a crucial role in determining the vacuum structure.

To support this interpretation, let us recall that the nonlocal Riegert action, when expanded in powers of the curvature, takes the form
\begin{equation}
S_{\text{anom}} = b \int \mathrm{d}^4 x\, \sqrt{g}\, \square R\, \Delta_4^{-1}\, \square R + \mathcal{O}(R^3),
\end{equation}
where \( \Delta_4 = \square^2 + \mathcal{O}(R) \) and \( b \) is a constant. Thus, up to quadratic order, the action becomes local and reduces to a term proportional to \( R^2 \), which is insufficient to capture vacuum polarization effects at large distances. Indeed, this local action yields an RSET proportional to the tensor \( H_{\mu\nu}^{(1)} \), which involves two derivatives of the curvature tensor and therefore decays as \( M R_0^2 / r^7 \). At large distances, this contribution is much smaller than the result we derived from the nonlocal effective action.

One might wonder whether the tachyonic instability discussed in the previous section could play a role in the results of Ref.~\cite{Boulware}. However, we have verified that no tachyonic instabilities arise in the particular case \( M = R_0/2 \) for conformal coupling \( \xi = 1/6 \). For these and other parameter choices that do not produce instabilities, we expect the Boulware vacuum to be perfectly well defined. Note that this is not necessarily inconsistent with a \( 1/r^3 \) decay, although it would be surprising if higher-order corrections dominated over the weak-field results in the large-distance limit.

A possible explanation, as already emphasized, is that the Weyl-invariant part of the effective action may substantially modify the asymptotic decay. This is precisely what occurs in the Schwarzschild geometry. Indeed, Ref.~\cite{Shapiro99} showed that the Riegert action produces an RSET that decays as \( M^2/r^6 \) in the Boulware vacuum, in agreement with earlier analytic approximations to the stress tensor~\cite{Anderson95}. However, subsequent numerical calculations revealed a decay of the form \( M/r^5 \)~\cite{AndersonPRL}, later confirmed through an analytical approach~\cite{universality}. Our results, combined with those  of Ref.\cite{Anderson2011}, suggest that a similar behavior occurs for the Bardeen geometry.

 \section{Discussion}\label{sec6}
For regular, horizonless spacetimes, the RSET can be evaluated in the weak-field regime using the nonlocal effective action derived via covariant perturbation theory \cite{nonlocal}. When the classical stress tensor sourcing the geometry is localized, the leading-order term scales as $M/r^5$, and depends solely on the total mass, while subleading corrections encode information about the star’s internal structure \cite{satz}. For nonlocalized sources, as in the Bardeen-type spacetime, the calculation presented in this work yields the same asymptotic behavior, consistent with the general results of Ref.~\cite{Anderson2011}.

Even in the weak-field limit, the RSET exhibits noteworthy features. Unlike idealized stellar models where the RSET often diverges at the surface, the Bardeen metric leads to a regular stress-energy tensor everywhere. Moreover, around $r\simeq R_0$, the energy density displays a rich structure, including several sign changes. This behavior differs between conformal and nonconformal fields, both in sign structure and in the amplitude of vacuum fluctuations near the origin.

A particularly interesting aspect of the Bardeen geometry, highlighted in this work, is the existence of tachyonic instabilities in the strong-field regime. In such cases, vacuum polarization effects may grow exponentially with time \cite{awakening}, indicating that the Boulware vacuum becomes unstable and thus physically inappropriate.

Another aspect that has been emphasized in this work is that  results obtained for the RSET using the nonlocal effective action in the weak-field regime differs from those obtained from the anomaly induced effective action, as happens for the Schwarzschild geometry. Therefore, the Weyl-invariant part of the effective action seems to be crucial to capture the correct behavior, at least for weak gravitational fields.

Future work will address the detailed behavior of the RSET in time-dependent weak gravitational fields, with particular focus on its nonlocal features and its asymptotic behavior at large $r$. The known sensitivity of the imaginary part of the effective action to the internal structure of oscillating stars \cite{Ddim}, as well as the sensitivity of the RSET to the details of the gravitational collapse \cite{Calmet}, strongly suggests a rich interplay between spacetime dynamics and quantum effects even in the weak-field regime. Work in this direction is in progress.

\section* {Acknowledgments} 
This research was supported by  Consejo Nacional de Investigaciones Científicas y Técnicas (CONICET).

\section* {Data availability}
The data are not publicly available. The data are available from the authors upon reasonable request.


\end{document}